# Diffraction 2006: Theoretical Summary

**Jeffrey R. Forshaw**
*School of Physics & Astronomy*
*University of Manchester, Manchester M13 9PL, UK.*
*E-mail:* `forshaw@mail.cern.ch`

This paper presents a summary of the theoretical presentations to the international workshop "Diffraction 2006". The range of topics covered during the workshop was quite broad and this summary is therefore somewhat selective covering recent developments in BFKL physics, exclusive processes, saturation dynamics, DIS and structure functions and coherence in QCD.

# 1. Developments in BFKL physics

BFKL physics is concerned with understanding QCD in the Regge limit of high centre-of-mass energy within the framework of perturbation theory[1]. It is the case that Regge-like behaviour does emerge out of QCD, giving the hope that we can understand objects like the pomeron using non-abelian quantum gauge field theory. The original leading order (LO) BFKL calculations of the 1970's led to the construction of an integral equation which describes gluon-gluon scattering in the Regge limit [1,2]. These calculations were followed many years later by the calculation of the NLO corrections to the integral equation [3,4]. More recently, effort has been directed at the problem of taking the solution to the integral equation and convoluting it with appropriate "impact factors" (also computed at NLO accuracy) in order to produce the first complete NLO BFKL calculations at the cross-section level. At the same time, the mathematical structure of high energy QCD has been the subject of closer scrutiny. Properties such as integrability and maximal transcendality have encouraged theorists to look more closely at the theory with potentially exciting consequences; the latest news was presented in the talk by Lipatov.

## 1.1 Integrability and maximal transcendality

It is possible to write the BFKL equation as a Schrödinger equation [5,6]:

$$H_{12}\psi(\vec{\rho}_1,\vec{\rho}_2) = E\psi(\vec{\rho}_1,\vec{\rho}_2)$$

where $\vec{\rho}_1$ and $\vec{\rho}_2$ represent the transverse positions of the reggeized gluons and the Hamiltonian is

$$H_{12} = \ln|p_1 p_2|^2 + \frac{1}{p_1 p_2^*}\ln(|\rho_{12}|^2) p_1 p_2^* + \frac{1}{p_2 p_1^*}\ln(|\rho_{12}|^2) p_2 p_1^* - 4\psi(1)$$

where $\rho_{12} = \rho_1 - \rho_2$. The co-ordinates and momenta are defined such that $\vec{\rho}_1 = (\rho_{1x}, \rho_{1y})$, $\rho_1 = \rho_{1x} + i\rho_{1y}$ and $p_1 = i\partial/\partial\rho_1$ etc.

Of particular interest are the energy eigenvalues since they locate the *j*-plane poles in scattering amplitudes arising as a result of vacuum exchange in the *t*-channel. In particular, the ground state energy determines the right-most singularity (i.e. the intercept of the BFKL pomeron).

The dimensionality of the problem can be reduced as a result of the fact that the Hamiltonian can be written as a sum of two parts, i.e. $H_{12} = \dfrac{h_{12}^* + h_{12}}{2}$ where

$$h_{12} = \ln|p_1 p_2|^2 + \frac{1}{p_1}\ln(\rho_{12})p_1 + \frac{1}{p_2}\ln(\rho_{12})p_2 - 2\psi(1).$$

---

[1] Although recent progress aims at pushing also into the strong coupling limit.

This holomorphic separability applies also to the case of any number of exchanged reggeized gluons interacting pairwise with nearest neighbours. In this case we recover the BKP equation:

$$H = -\sum_{k<l} \frac{T_k \cdot T_l}{N_c} H_{kl} .$$

The BKP equation was originally derived as far back as 1980 as the "minimal way" to unitarize BFKL (in the leading $N_c$ approximation) [7,8].

Remarkably, the general solution to this Schrödinger equation is accessible. The key is to realise that the problem is identical to that of a Heisenberg XXX spin chain where the spins are the generators of the group SL(2,C) and that problem is known to be integrable (which, roughly speaking, means there are as many operators which commute with the Hamiltonian as there are degrees of freedom) [9,10]. At first sight, the problem then reduces to identifying the ground state wavefunction and an appropriate raising operator. However, the lack of a "highest weight" state meant that a more sophisticated approach was needed and progress was eventually made in [11] where, amongst other things, it was shown that the solution for three reggeons gives the odderon solution with intercept equal to unity. See also [12] and references therein.

Conformal symmetry is crucial to the integrability of the BKP equation and that is lost in QCD beyond leading order (e.g. when the QCD coupling starts to run). However, it is not lost in N=4 supersymmetric QCD and now the Hamiltonian remains holomorphically separable beyond LO. Moreover, it also appears to be built only using harmonic sums (i.e. $\psi$ functions and their derivatives). This suggests a "principle of maximal transcendality"; a principle which is supported by the fact that the N=4 DGLAP splitting functions at two-loops (which can be computed by traditional means) appear also to be maximally transcendental [13]. More specifically (in *j*-space) they can always be built using harmonic sums like

$$S_a(j) = \sum_{m=1}^{j} \frac{1}{m^a} \quad , \quad S_{abc...}(j) = \sum_{m=1}^{j} \frac{1}{m^a} S_{bc...}(m) \quad .$$

What are we to make of this principle? Well, Kotikov, Lipatov, Onischenko & Velizhanin recently invoked it in order to predict the three-loop splitting function in N=4 QCD [14] by extracting the maximally transcendental parts of the three-loop calculations by Moch, Vermaseren & Vogt in ordinary QCD [15]. Unfortunately it is not easy to check their result (the N=4 QCD direct calculation is even harder than in QCD) but it has been shown to be correct in the $j=1$ and $j=\infty$ limits. Eden & Staudacher have also used the principle to make a prediction for the all-loops splitting function at large $j$. The significance of this underlying mathematical structure remains to be established but it does appear to be yet another hint pointing to an important connection between QCD and string theory.

## 1.2  NLO BFKL

For several years now we have known the BFKL equation at NLO for forward scattering, i.e. $gg \to gg$ at $t = 0$ [3,4]. To compute the NLO corrections to non-forward scattering requires more work and Fadin reported on the latest progress towards that goal. He pointed out that the "two-gluon" contribution illustrated in Fig.1 was the stumbling block which has now been surmounted [17]. In addition, first (partial) results were presented on the representation of the kernel in transverse co-ordinate space. That should prove useful both for checking the conformal structure and also in understanding the dipole formalism at NLO.

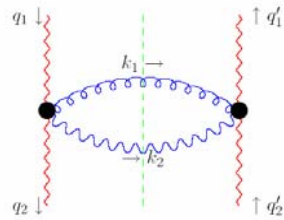

*Figure 1: The troublesome two-gluon contribution*

Apart from knowing the gluon scattering amplitude we must also compute the coupling of the gluons to the external particles if we are eventually to compute observable cross-sections and these too must be evaluated at NLO. Chachamis told us that the calculation of the NLO photon impact factor, illustrated in Fig.2, is more-or-less completed and that it will soon be possible to compute the $\gamma^*\gamma^* \to$ hadrons total cross-section at NLO [18,19]. He showed explicit numerical results for the impact factor and a comparison to the LO result. Chachamis also pointed out that the results will however remain sensitive to how the QCD coupling is made to run (since it is still only one-loop running).

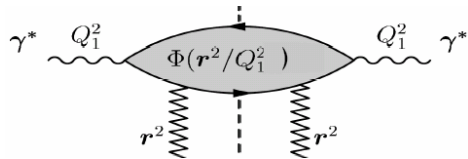

*Figure 2: The photon impact factor*

The $\gamma^*\gamma^* \to$ hadrons total cross-section will not be the first completed BFKL calculation at NLO for, as Papa told us, the NLO calculation of $\gamma^*\gamma^* \to \rho\rho$ (at $t=0$) has already been completed [20,21]. The relevant impact factor is illustrated in Fig.3 where $\phi_\parallel(z)$ is the leading twist-2 meson distribution amplitude. Apparently, there is evidence for large and negative NLO corrections and there is quite some sensitivity to the choice of renormalization scale with a hint (using PMS) that the appropriate scale choice is considerably larger than the natural value corresponding to the photon virtuality.

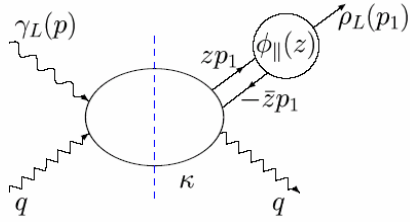

*Figure 3: The impact factor for rho meson production*

Apart from exclusive processes, one can also try and learn more about NLO BFKL through inclusive jet production and that subject was the focus of Schwennsen's talk. The goal here is to compute the cross-section for single jet production in scattering processes at high centre-of-mass energy. At LO this is fairly straightforward however at NLO one needs the real emission vertex at NLO and that includes the two terms shown schematically in Fig.4. One needs to "break open" the integral over the outgoing real partons (an integral which is always performed when computing the BFKL kernel for exclusive processes) and implement a jet algorithm [22]. The jet vertex has been computed and the goal now is to compare to experiment and also to the predictions based upon the more traditional collinear factorization approach.

$$\mathcal{K}_r \sim \rhd\!\!- \;+ \int \!\lhd$$

*Figure 4: The real emission vertex at NLO*

On the same lines, Sabio Vera discussed the possibility to look at the azimuthal correlations between the two widely seperated jets in dijet events (or between the electron and jet in DIS). In particular he emphasised that quantities such as $<\cos(m\phi)>$ should yield information on the role played by higher conformal spin. He presented results including some NLO effects and showed that NLO effects are important in bringing about much less decorrelation than LO, in agreement with the data [23].

## 2. Exclusive processes

### 2.1 Electroproduction

In diffractive electroproduction we received several presentations: one from Igor Ivanov on the production of spin-3 mesons and one from Goloskokov on vector meson production. Shaw also presented results on diffractive electroproduction of vector mesons and on DVCS, and Goncalves talked about diffractive electroproduction off nuclei; we will discuss these in Section 4.

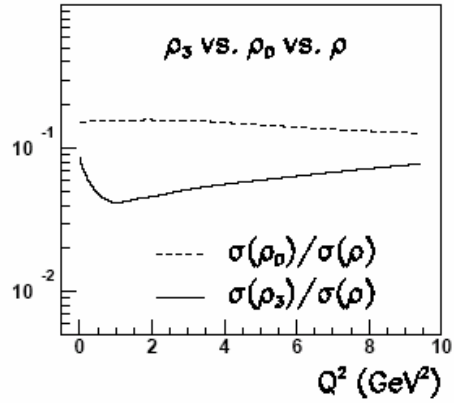

*Figure 5: Spin-3 and D-wave meson production rates relative to the case of the rho.*

Within the framework of $k_T$-factorization Igor Ivanov presented results on D-wave ($\rho''(1700)$) and on spin-3 ($\rho_3(1690)$) meson production in diffraction. These two mesons are spin-orbital partners but the $\rho_3$ is absent from the incoming photon wavefunction and its production is therefore essentially off-diagonal. He presented a prediction that both processes should occur at non-negligible rates, as illustrated in Fig.5 [24,25]. Moreover, the relative contributions arising from production using longitudinal photons is very different for the two cases and each is very different from $\rho$ meson production. In particular the ratio

$$R_{LT} = \frac{\sigma_L}{\sigma_T}\frac{m_V^2}{Q^2}$$

is, at $Q^2 = m_V^2$, equal to 1 for the $\rho$ whilst it is 9/289 for the $\rho''$ and 6 for the $\rho_3$. The largeness of the ratio for $\rho_3$ implies that even at quite low $Q^2$ its production may be dominated by a helicity violating amplitude. Calculations also indicate that $\rho_3$ production should be dominated by production via large size dipoles (~2fm), in which case one might look here for an enhancement of saturation dynamics. All that remains now is for the experimenters to uncover a sample of spin-3 mesons produced in electroproduction and Ivanov informed us that investigations are well underway.

Staying on the theme of diffractive electroproduction, Goloskokov presented results for $\gamma^* p \to \rho p$, including production using transverse photons, within the formalism of Generalized Parton Distribution functions (GPDs). This approach has formerly been used mainly for production using longitudinally polarized photons where it is clear that the use of QCD perturbation theory is justified and one can use such processes to constrain the GPDs. Extending the formalism to include also transverse photons is performed via the inclusion of Sudakov corrections and intrinsic transverse momentum in the meson's wavefunction [26]. It is argued that the inclusion of these effects makes a perturbative calculation viable and the comparison to data is impressive; the calculations agree with available HERA data including the wide range of spin-density matrix element measurements. Goloskokov also pointed out that an asymmetry measurement made using a longitudinally polarized beam and target (e.g. as could be performed at HERMES and COMPASS) would be sensitive to the spin of the gluon in the proton.

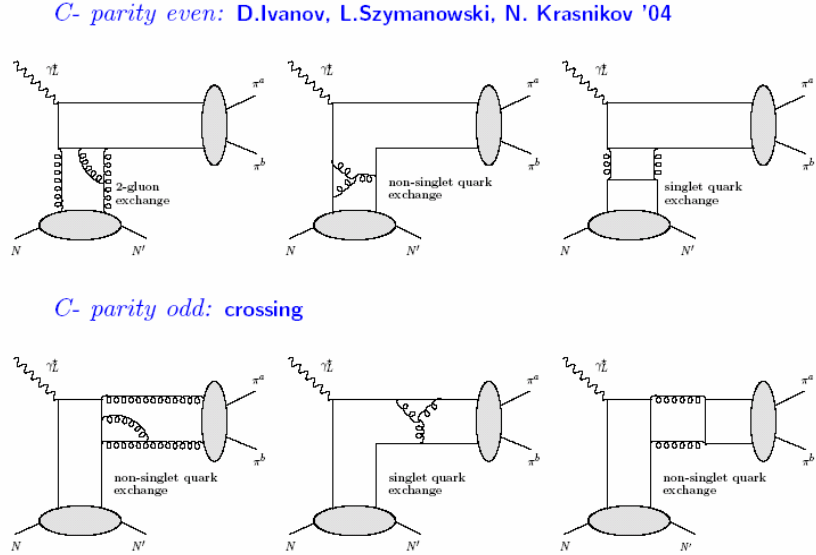

*Figure 6: NLO contributions to two-pion electroproduction.*

We can also gain access to information on GPDs from the (diffractive and non-diffractive) electroproduction of pion pairs in $\gamma^* N \to \pi\pi N$ and a calculation of this process was discussed by Dmitri Ivanov. A full NLO calculation was perfomed and typical graphs are illustrated in Fig.6 [27]. In Fig.7, the predictions are compared to data from HERMES at $<x>=0.16$, $<Q^2>=3.2(3.3)\,\text{GeV}^2$ and $<|t|>=0.43(0.29)\,\text{GeV}^2$ collected off hydrogen (deuterium) targets. One can readily see that the NLO corrections are not large. One should however be aware that the data are really at rather low values of $Q^2$ where the factorization between hard scattering, GPD and two-pion distribution amplitude might not be expected to hold so well. Moreover, the calculations are leading twist in the pion distribution amplitude and higher twist corrections should play a role at low enough $Q^2$.

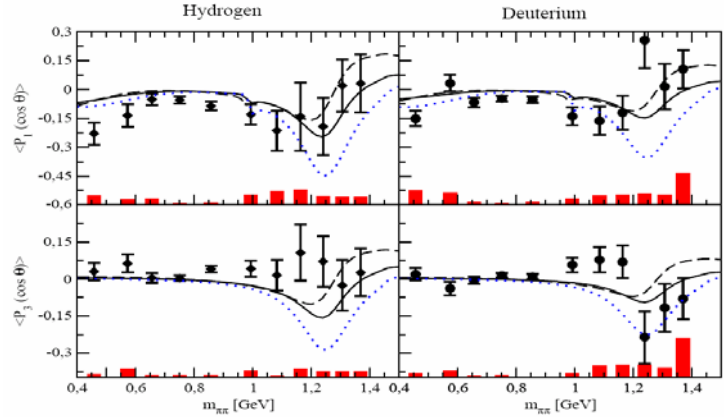

*Figure 7: Comparison of predictions for two-pion production to HERMES data. LO (dashed black line), NLO (solid black line). The dotted line is the calculation without account of 2-gluon exchange in the t-channel.*

The issue of how QCD factorization works in exclusive processes was explored further in Segond's talk where he specifically looked at the theoretical test-bed process $\gamma^*\gamma^* \to \rho_L \rho_L$ (the choice of longitudinal mesons avoids any potential problems associated with large end-point contributions) [28]. He specifically discussed the factorization properties of the scattering amplitude in two limits: (1) the limit $W^2 \ll Q^2$, in which the amplitude for production using transverse photons factorizes as in Fig.8 into a product of hard scattering, a generalized distribution amplitude and two meson distribution amplitudes and (2) the limit $Q_1^2 \gg Q_2^2$, in which the amplitude for production using longitudinal photons factorizes as in Fig.9 into a product of a hard scattering amplitude and a transition distribution amplitude.

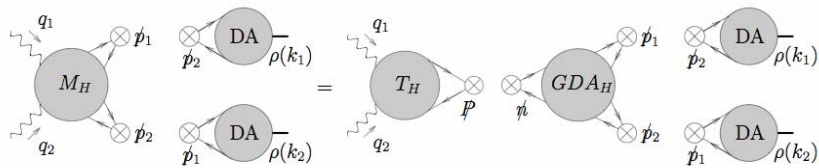

*Figure 8: Factorization for transverse photons at W << Q.*

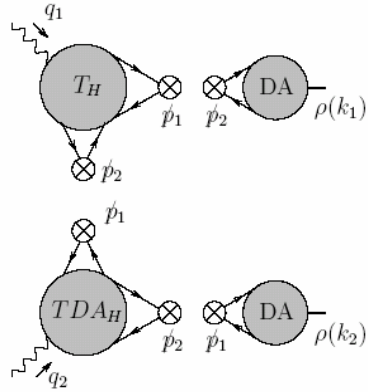

*Figure 9: Factorization for longitudinal photons in the DIS limit.*

## 2.2 Central production

There is growing interest, both experimentally and theoretically, in another type of exclusive process, namely $pp \to p + X + p$ at high energies where the protons scatter through small angles, losing only a small fraction of their longitudinal momentum [29,30]. The possibility to install detectors some 420m downstream from the interaction point in order to measure the outgoing protons' four-momenta [31] has opened up the possibility to make accurate measurements of the mass of the central system $X$ and interest has been particularly focussed upon the case where $X$ could be a single Higgs boson.

One important source of contamination to purely exclusive events arises when the central system of interest is produced in conjunction with additional particles. Royon stressed the need to understand this class of non-exclusive events and that uncertainties in the large $\beta$ behaviour of the gluon in the pomeron extracted using HERA data would in turn lead to uncertainties in central exclusive production (CEP) [32].

The question of "gap survival" is a thorny one in the theoretical studies of CEP. Eikonal models have been used to make predictions for the fraction of events which survive without additional radiation after accounting for soft re-scattering effects [33,34]. However these models are not based on fundamental theory. Nevertheless, they have had their successes in simultaneously describing HERA and Tevatron data, and the general properties of the so-called soft underlying event [35,36,37]. Royon pointed out that correlations between the azimuthal angles of the outgoing protons are expected to be sensitive to the model of gap survival and that measurements using the existing low angle detectors at DØ could be used to test theoretical models using Tevatron data.

A brief overview of the theoretical status of CEP was presented by Forshaw [30] who also presented predictions for expected rates in a variety of new physics scenarios. Apart from Standard Model Higgs production there is particular interest in Higgs production in the MSSM in the so-called intense coupling regime [38] or with explicit

CP violation in Higgs sector [39]. In both cases the production rates can be greatly enhanced and a measurement of the Higgs mass can be obtained in channels which could be very difficult to study using more conventional means. Forshaw also discussed the possibility to study stable gluino production (as might occur in "split supersymmetry") and measure the gluino mass, should it be lighter than ~350 GeV [40].

Apart from the production of new physics at high mass, the central system could be light, e.g. $X = \eta', \eta_c, \eta_b$. Of course production of such low masses is not perturbative and Szczurek presented his results using a non-perturbative model inspired by the Durham calculation which is used for larger masses, in conjunction with the $\gamma\gamma$ fusion process [41]. The model is unable to reproduce the WA102 data (at $W$ = 30 GeV) for $\eta'$ production, falling far below the data. Apart from the inherent uncertainties in the model, one should note that the calculation is performed using only gluon fusion and quark contributions may well be playing an important role at these energies.

### 3. Dipoles, saturation and unitarization

At high enough centre-of-mass energy it has long been anticipated that non-linear QCD dynamics will begin to play a role, in order to prevent the power-like growth of cross-sections with increasing energy from violating unitarity. However, the discovery of this non-linear gluon dynamics in the data has been elusive, certainly in the perturbative domain. Shaw presented the results of a phenomenological investigation to ascertain the extent to which saturation dynamics appear to be needed by existing data. His calculations are performed within the dipole model approach [42]. The strategy is to use the precise HERA data on the $F_2$ structure function to constrain the universal dipole cross-section and this is then used to make predictions for other observables, notably $F_2^c$, DVCS, exclusive vector meson production and the diffractive DIS structure function $F_2^{D(3)}$. Shaw concluded that the dipole formalism works very well indeed, being able to accommodate all of the electro/photo-production data. It is certainly true that models for the dipole cross-section which contain saturation effects do very well in explaining the data. However, it is also true that all of the data, with the exception of $F_2$ data at low $Q^2$ (i.e. $< 2$ GeV$^2$), can also be explained using a dipole model which does not include any saturation. It appears that the data are unfortunately at insufficiently low $x$ to probe saturation dynamics in a clearly perturbative domain.

Over the past few years there has been perhaps a resurgence of interest in non-linear evolution in QCD. Hatta presented us with a summary of some of the latest developments in the field [43]. The situation is illustrated in Fig.10 where one defines a saturation scale $Q_s(Y)$ below which non-linear effects play an important role.

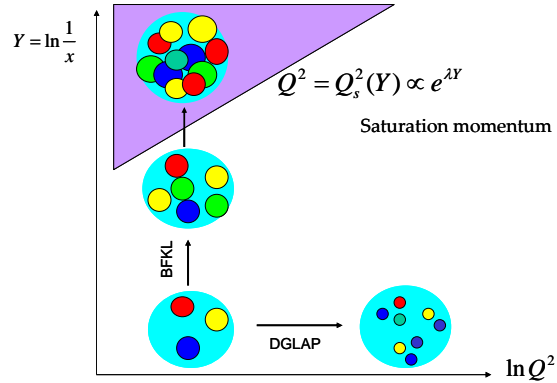

*Figure 10: Illustrating the non-linear regime at small-x*

According to the wisdom of the BK equation [44,45] (and the more sophisticated JIMWLK approach [44,46-51]), the amplitude $T(r,Y)$ for scattering a dipole of size *r* off some target evolves in *Y* as a travelling wave; a phenomenon known as "geometric scaling" and illustrated in Fig.11.

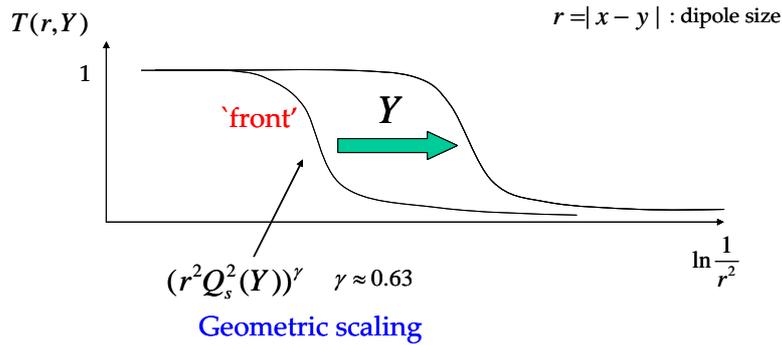

*Figure 11: The dipole amplitude and geometric scaling*

The position of the front of the travelling wave is fixed by the saturation scale, $Q_s(Y)$. One consequence of this behaviour is that diffractive DIS (DDIS) should be dominated by dipoles of size $r \approx 1/Q_s < 1/\Lambda_{QCD}$. However, Hatta pointed out that important "pomeron loop" contributions are missing from the BK/JIMWLK approach and that including them changes the physics quite dramatically. The position of the travelling wave-front now becomes a stochastic variable, as illustrated in Fig.12, and the observed scattering amplitude is obtained by averaging, i.e.

$$<T(r,Y)> = \int dQ_s \, P(Y,Q_s) T(rQ_s).$$

Now upward fluctuations in the saturation scale mean that saturation physics can enter at higher values of $Q^2$ with the consequence that DDIS becomes dominated by the scattering of small size dipoles at sufficiently large *Y* [43].

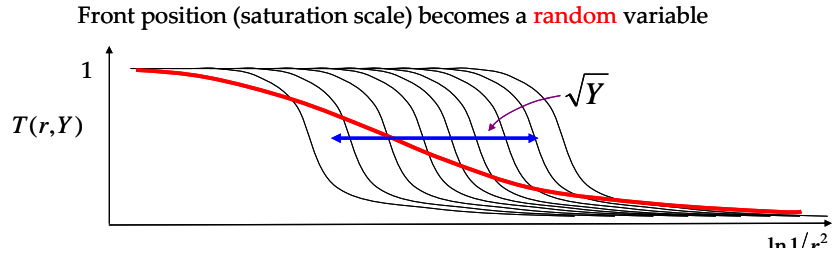

*Figure 12 : Pomeron loops mean that the saturation scale a random variable*

Perhaps one of the most promising avenues on which to search for saturation dynamics is to look in electroproduction off nuclei and Goncalves presented results of a preliminary study into the possible role of saturation dynamics in DDIS off nuclear targets, as might be probed at a future eRHIC or THERA experiment [52].

Remaining on the topic of collisions involving nuclear targets, Gay-Ducati presented predictions for di-lepton production in the forward and backward directions in proton-nucleus interactions [53,54]. She emphasised that di-leptons produced at forward rapidities (i.e. in the direction of the incoming proton) constitute a good probe of saturation dynamics whilst di-lepton production at backward rapidities should be sensitive to large-$x$ nuclear effects. The idea is that di-leptons, being colour neutral offer a cleaner probe of interesting physics than, e.g. hadron production.

Schäfer also presented the results of a study into the role of non-linear effects arising in the interaction of a small probe with a nuclear target. Crudely stated, the idea is to generalize the Glauber-Gribov approach to account more completely for the colour dynamics of QCD [55].

## 4. DIS and structure functions

Corcella reminded us of the need to properly account for soft gluon effects at large $x$ and he presented the results of a toy model which illustrated the possible impact of re-summation effects. He noted a visible impact (10-20%) upon the up-quark distribution in the range $0.55 < x < 0.8$ [56]. It will be interesting to see how much this correction is after evolving using the NNLO evolution equations.

Zoller made the interesting observation that one expects much more nuclear shadowing in charged current DIS if the probe is a left-handed positively charged W boson since in that case one is sensitive to quark dipoles of a size equal to the inverse of the strange quark mass [57]. In contrast, right-handed W bosons scatter using dipoles whose size is determined by the inverse of the charm quark mass. One way to probe this strong L-R asymmetry is through the quantity

$$\Delta(xF_3) = xF_3^{\nu} - xF_3^{\bar{\nu}}.$$

For example, in DIS off nuclei one should measure the quantity

$$R = \frac{\delta(\Delta x F_3)}{A \Delta x F_3} = \frac{\delta\sigma^A_L - \delta\sigma^A_R}{A(\sigma_L - \sigma_R)} \quad \text{and} \quad \delta\sigma^A_\lambda = A\sigma_\lambda - \sigma^A_\lambda.$$

The low-$x$ behaviour of the spin structure functions was the subject of a talk by Greco. He presented results of a complete resummation of all leading logarithms (including double and single logarithms in $x$). At low $Q^2$ and low $x$ it is argued that the structure function $g_1$ does not much depend upon $x$ and that it can be close to zero in the low $x$ range now probed by COMPASS. He also suggested that the experimenters look at the data as a function of the invariant $2p \cdot q$ since it is expected that $g_1$ turns negative at sufficiently large values and that this point would be sensitive to the relative balance between the quark and gluon densities.

## 5.     Coherence in QCD

Colour coherence is generally believed to be an essential property of QCD. In essence the idea is that soft gluons emitted at large angles cannot resolve emissions at low angles, as illustrated in Fig.13.

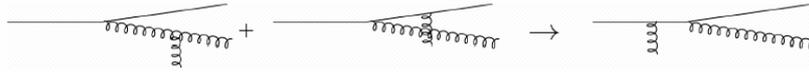

*Figure 13: Soft gluon emission at wide angles does not resolve emission at lower angles.*

In fully inclusive observables the effects of soft gluons are known to cancel. However for less inclusive quantities, where there is some restriction on the phase space of emitted gluons, large logarithms can arise and generally need to be resummed. Kyrieleis presented results which indicate that coherence in QCD may not hold in some instances [58]. In particular he focussed upon the "gaps-between-jets" observable in which at least two jets are produced and there is a requirement that no further jets with transverse momentum above a scale $Q_0$ be produced in between them. For this observable, he demonstrated the possibility to produce super-leading logarithms due to soft gluons emitted collinear with an incoming parton. Ordinarily coherence would guarantee that these contributions would cancel and it would be safe to absorb all collinear logarithms into the incoming parton distribution functions. Non-cancellation is tantamount to a breakdown of the "plus prescription" in the evolution of the incoming parton distribution function for scales above the scale $Q_0$ which defines the gap. The existence of these super-leading logarithms (which occur for the first time at a high order in perturbation theory) and the consequent breakdown of coherence is so far only suggested by [58] and more work remains to be done before one can be certain.

**Acknowledgements**

My sincerely thanks go to the organizers of a splendid workshop and to the many speakers whose presentations I enjoyed so much. Thanks are due to everyone whose slides and figures I have used in producing this talk and apologies are due to those whose contributions did not make it into this final version.